\documentclass[twocolumn,showpacs,preprintnumbers,amsmath,amssymb,
 aps,
 prb,
 lengthcheck,
]{revtex4-1}
\usepackage{color}
\usepackage{graphicx}
\usepackage{dcolumn}
\usepackage{bm}
\usepackage{hyperref}
\usepackage{makeidx}
\makeindex
\def\>{\right\rangle}
\def\<{\left\langle}
\def\be{\begin{equation}}
\def\ee{\end{equation}}
\def\ba{\begin{array}{ccc}}
\def\ea{\end{array}}

\def\beq{\begin{eqnarray}}
\def\eeq{\end{eqnarray}}

\def\bpsi{\mbox{\boldmath $\psi$}}

\begin{document}

\preprint{APS/123-QED}
\title{Environmental induced renormalization effects in quantum Hall edge states}
\author{A. Braggio$^{1}$, D. Ferraro$^{1,2,3}$, M. Carrega$^4$, N. Magnoli$^{2,3}$, M. Sassetti$^{1,2}$}
 \affiliation{
$^1$ CNR-SPIN, Via Dodecaneso 33, 16146, Genova, Italy.\\ 
 $^2$ Dipartimento di Fisica, Universit\`a di Genova,Via Dodecaneso 33, 16146, Genova, Italy.\\
$^3$ INFN, Sezione di Genova, Via Dodecaneso 33, 16146, Genova, Italy.\\
$^4$ NEST, Istituto Nanoscienze - CNR and Scuola Normale Superiore, I-56126 Pisa, Italy.}
 \date{\today}

\begin{abstract}
We propose a general mechanism for renormalization of the tunneling exponents in edge states of the fractional quantum Hall effect.
Mutual effects of the coupling with out-of-equilibrium $1/f$ noise and dissipation are considered both for 
the Laughlin sequence and for composite co- and counter-propagating edge states with Abelian or non-Abelian statistics.
For states with counter-propagating modes we demonstrate the robustness of the proposed mechanism in the so called disorder-dominated phase. Prototypes of these states, such as $\nu=2/3$ and $\nu=5/2$, are discussed in detail and the rich phenomenology induced by the presence of a noisy environment is presented. The proposed mechanism justifies the strong renormalizations reported in many experimental observations carried out at low temperatures.  We show how environmental effects could affect the relevance of the tunneling excitations, leading to important implications in particular for the $\nu=5/2$ case.  
 \end{abstract}

\maketitle

\section{Introduction}
States living on the boundary of Fractional Quantum Hall (FQH) systems represent one of the more intriguing examples of 
one-dimensional interacting electron gas.\cite{DasSarma} The general theory describing these edge states involves the idea of Chiral Luttinger 
Liquid ($\chi$LL).\cite{Wen95, Wenbook}
In particular, for the simple Laughlin sequence \cite{Laughlin83}  at filling factor $\nu=1/(2n+1)$, 
with $n \in \mathbb{N}$, all the properties of the system,  including the fractional charge and statistics of edge excitations, are described  
in terms of a single chiral bosonic field.  More involved is the description of states belonging to the Jain sequence \cite{Jain89} at 
filling factor $\nu = p/ (2n p + 1)$, being $n \in \mathbb{N}$ and $p \in \mathbb{Z}$, where the introduction of a charge bosonic field and 
additional neutral bosonic modes are required by the proposed hierarchical theories leading to an hidden SU($|p|$) 
symmetry.\cite{Wen91,Kane95}
Recently, great interest was dedicated to more exotic states, such as \cite{Willett87} $\nu=5/2$ where different models were proposed, 
with excitations supporting both Abelian\cite{Halperin83,Halperin93,Wen90} or 
non-Abelian\cite{Moore91,Wen91b, Fendley07, Levin07, Lee07, Bishara08, Carrega12} statistics. To date, different experimental 
observations\cite{Radu08,Bid10, Dolev08, Venkatachalam11} suggest the non-Abelian anti-Pfaffian model as the proper 
candidate for $\nu=5/2$ even if the debate is still open.\cite{Lin12} 
 In the effective field theories the peculiar non-Abelian properties are encoded in an additional conformal field, which belongs to the 
 Ising sector.\cite{ Nayak08}
 The non-Abelian nature of the 
excitations of this state arose interest in perspective of possible applications to topologically protected quantum computation.\cite{Nayak08}
The simpler experimental test for all these models is the study of transport properties in a quantum point contact (QPC) geometry.\cite{Chang03} 
In absence of interactions between edges and external degrees of freedom, the power-law behavior 
of the transport properties in the QPC geometry, as a function of bias or temperature, directly reflects the universal exponents of the $\chi$LL theory.

Unfortunately, sometimes strong discrepancies between the predictions of such theories and experimental observations are reported. For example, 
even for the simple Laughlin sequence \cite{Roddaro03, Roddaro04} the behavior of the differential conductance, as a function of the voltage, is in 
qualitative agreement with predictions only at high temperature showing a peak at zero bias. However, decreasing temperature, the observed peak 
turns into a completely unexpected dip.

 Anomalous current/voltage characteristics have been also measured for other filling factors such as  $\nu=2/5$ in the Jain sequence.\cite{Chung03}
Furthermore, renormalizations of the $\chi$LL exponents are sometimes crucial to fully explain the measured crossover of the tunneling 
charges at low temperatures.\cite{Chung03, Bid09, Dolev10, Ferraro08, Ferraro10a, Ferraro10b, Ferraro10c, Carrega11}

Possible explanations for these disagreements have been traced back to the inhomogeneity of the filling factor below the QPC due to the action 
of the electrostatic gates\cite{Roddaro04, Lal08}, or to an energy dependent tunneling amplitude caused by the extended nature of the 
contact.\cite{Overbosch09, Chevallier10}
 Alternatively, various mechanisms leading to the renormalization 
of the Luttinger parameters through coupling with external environments have been proposed. They range from
the coupling with one dimensional phonons \cite{Rosenow02, Khlebnikov06}, edge reconstruction induced by the smoothness of the confinement 
potential \cite{Yang03}, possible Coulomb interaction between the different edges \cite{Mandal02, Papa04}, to interactions  with a compressible 
component of a composite fermions liquid with very small longitudinal conductivity.\cite{Shytov98,Levitov01}

Many of these approaches have focused on the Laughlin case and cannot be easily extended to composite edge states 
 where anomalous behaviors are usually observed.
In particular, many of the above mechanisms  are not robust against the disorder induced intra-edge electron tunneling, an unavoidable effect in real 
samples responsible for the equilibration of the different channels. This is a crucial ingredient in explaining the universal quantization of the 
conductance in presence of counter-propagating modes.\cite{Kane94, Kane95, Levin07, Lee07, Bishara08, Carrega12}

Recently, Dalla Torre \emph{et al.}\cite{DallaTorre10, DallaTorre11} observed that the interplay between the $1/f$ noise, generated by 
the external environment, and the dissipation induced by the cooling setup 
could lead to the renormalization of the Luttinger parameter for one dimensional systems of cold atoms.
 
In this paper we will apply this idea to the case of the edge states in the FQH  effect, taking in account the peculiar chiral 
nature of the $\chi$LL theories and investigating the effects of an external noisy environment.  The $1/f$ noise, a 
quite universal and unavoidable perturbation in any electronic circuitry, can be indeed generated by trapped 
charges in the semiconductor substrate.\cite{Paladino02, Muller06}
These sources of noise, with $1/f$ spectrum, drive stochastically the system into an out-of-equilibrium condition and the stationary condition is 
recovered, by the dissipation mechanism. 
For one dimensional electron systems and $\chi$LL, has been considered in literature different mechanisms: from the coupling 
with metallic gates used to confine 
the electron gas \cite{Cazalilla06} to the coupling with electromagnetic environment \cite{Sassetti94,CastroNeto97,Safi04} or with other 
systems.\cite{Shytov98, Levitov01}
In this paper we will discuss the consequences caused by the joint presence of $1/f$ noise and dissipation mainly because 
those effects are unavoidable in order to get a realistic description of the physical systems.

A relevant advantage of the proposed mechanism 
is its robustness to the disorder dominated phase, a key feature in order to apply the model to edge states with counter-propagating channels, 
such as $\nu=2/3$ and $\nu=5/2$. The aim of this paper is to present a detailed analysis of this fact and its applicability to real cases.

The paper is organized as follow. In Sec.\ref{laughlin} we consider the Laughlin sequence. 
Using this paradigmatic example we introduce the notations and the general methods that we will use later for the composite edge case, which is the 
main issue of the paper. The effects of the renormalization induced by the noisy environments and the possible 
consequences on the QPC transport are discussed. In Sec.\ref{Jainseq} we analyze the effect of the noisy environment on the Jain 
sequence limiting for simplicity only to the two-modes cases $\nu=2/5,2/3$. In particular for the co-propagating case $\nu=2/5$ 
we investigate how the scaling becomes non-universal and also dependent on the strength of the intra-edge Coulomb interaction 
when the external noise is present. This is strongly different from the standard hierarchical result where the scaling 
dimension is predicted to be independent from interaction between the modes. 
Discussing the case of counter-propagating modes (i.e. $\nu=2/3$), we exploit this condition to get the disorder dominated phase in presence of a 
noisy environment.
In Sec.\ref{fivehalf} 
the properties of the anti-Pfaffian model for $\nu=5/2$ 
as a function of the strength of the $1/f$ noise are analyzed, showing the regions of the parameters space where the 
elementary excitation with non-Abelian nature could dominate. We finally discuss the counterintuitive result that the external noise 
could help the manipulation of non-Abelian excitation in the QPC geometry. Conclusions are summarized in Sec.\ref{Conclusions}.    

\section{Laughlin sequence}
\label{laughlin}

Let us consider the edge states of a quantum Hall fluid, described in terms of $\chi$LL theories, and investigate the effect induced by the 
joint presence of an external environment, $1/f$ noise and dissipation. We stress that due to the presence of $1/f$ noise we have to face 
with an out-of-equilibrium problem, therefore in the following we will employ proper techniques, i.e. the Keldysh contour formalism.\cite{Schwinger61,Keldysh64, Rammerbook, Kamenev09}

\subsection{Model}
We start our analysis considering the Laughlin sequence \cite{Laughlin83} with filling factor $\nu=1/(2n+1)$, being $n \in \mathbb{N}$.
The Lagrangian density of the $\chi$LL for an infinite edge is described in terms of a single bosonic mode
\be
\label{free_Laughlin}
\mathcal{L}_0=\frac{1}{4 \pi \nu}  \partial_{x}\varphi  ( -\partial_{t} -v \partial_{x} ) \varphi,
\ee
where $\varphi$ is a right-moving field along the edge with propagation velocity $v$. 
 
In view of dealing with an out-of-equilibrium system, we treat the problem in the Keldysh contour 
formalism. According to the standard path integral 
formulation, the non-equilibrium problem 
on the doubled time contour is easily encoded 
by introducing the bosonic field in the forward/backward time branch $\varphi^{\mathrm{f}/\mathrm{b}}$. 
We refer the reader to Ref.~\onlinecite{Rammerbook} and 
Ref.~\onlinecite{Kamenev09} for the general treatment of these issues, and  to Ref.~\onlinecite{Martin05} for the applications of these 
methods to the edge states of the quantum Hall effect.

It is useful to write $\varphi^{\mathrm{f}/\mathrm{b}}= (\varphi^{\mathrm{cl}}\pm \varphi^{\mathrm{q}})/\sqrt{2}$ where 
 $\varphi ^\mathrm{cl}$ ($\varphi ^\mathrm{q}$) represents the so called classical (quantum) component of the field.\cite{Kamenev09} In terms of 
 the classical-quantum basis the bosonic Green's functions (GFs) are enclosed in the matrix 
 \be
\mathcal{G}_{a,b}(x, t)=-i\langle \varphi^{a}(x,t)\varphi^{b}(0,0)\rangle=\left(
 \ba
 G^{K}(x, t)& G^{R}(x, t)\\
 G^{A}(x, t)& 0\\
 \ea
 \right),
 \label{KeldyshGF}
 \ee
 where $a, b=\mathrm{cl, q}$. With $G^{R}$, $G^{A}$ and $G^{K}$ 
 the retarded, advanced and Keldysh GFs respectively.\cite{Kamenev09}
In terms of these fields and in Fourier transform, defined as 
$\varphi^{\mathrm{cl/q}} (q, \omega)=\int dx dt\  e^{i(\omega t-q x)}\  \varphi^{\mathrm{cl/q}} (x,t)$, the free Keldysh action, 
deduced from Eq. (\ref{free_Laughlin}), reads

\be
\label{KeldyshAction}
\mathcal{S}_{0}= \frac{1}{2}\sum_{q\neq0,\omega}\   
(\Phi^*(q, \omega))^T\cdot \mathcal{G}_{0}^{-1}(q, \omega)\cdot\Phi(q, \omega),
\ee
with the vector $\Phi=(\varphi^{\mathrm{cl}}, \varphi^{\mathrm{q}})^T$. 
The matrix kernel of the action 
is\cite{Kamenev09} 
\be
\label{chiLLGF}
\mathcal{G}^{-1}_{0}(q, \omega)=\frac{q}{2\pi\nu}\left[
\begin{array}{cccc}
0&(\omega-i\epsilon)- vq\\
(\omega+i\epsilon)- vq & 2iq\epsilon \mathrm{sgn}(\omega)
\end{array}
\right],
\ee
where $\epsilon\to0$ is the standard regularization factor\cite{Note1} and the top-left $0$ component corresponds to the standard 
continuum limit of the 
Keldysh action.\cite{Kamenev09} Inverting the Kernel matrix and taking the cl-q component $(\mathcal{G}_{0})_{\mathrm{cl},\mathrm{q}}$
we get, according to Eq. (\ref{KeldyshGF}), the retarded GF for the free bosonic fields 
\be
\label{ChiralRetardedGF}
G_{0}^{\mathrm{R}}(q, \omega)=\left(\frac{2\pi\nu}{q}\right)\frac{1}{(\omega+i\epsilon)- vq}
\ee
and analogously for the advanced GF, $G_{0}^{\mathrm{A}}=(\mathcal{G}_{0})_{\mathrm{q},\mathrm{cl}}$. From the linear response theory 
one can show that the current along the Hall bar is given by $I=\nu \mathfrak{g}_0 V_H$ 
with $V_H$ the Hall potential and $\mathfrak{g}_0=e^2/h$ the quantum of conductance.\cite{Kane95}

We discuss now the influence
of $1/f$ noise term at low energies.\cite{Paladino02, Muller06} This contribution can be described in terms of a classical stochastic external 
potential $f(x,t)$ 
that describes the effective interaction of the edge with the localized trapped 
charges. The correlation function of the external force   
is $K(q, \omega)=\langle f^{*}(q, \omega) f(q, \omega) \rangle=F/|\omega |$ where $F$ is the strength of the noise. For simplicity the noise 
is assumed $\delta$-correlated in space, a natural assumption for short-range impurities in the low energy/long wavelength limit. Note 
that the presence of this time dependent external force brings the system out-of-equilibrium.

The external gaussian random force $f(x,t)$ couples directly with the electron density 
$\rho= \partial_{x}\varphi/(2\pi)$ 
of the $\chi$LL.
In the Keldysh formalism this interaction is  \cite{Note1b}
\be
\mathcal{S}_{f,\varphi}=\!\!\int\!\! dt \left(\mathcal{L}_{f,\varphi^{\mathrm{f}}}-\mathcal{L}_{f,\varphi^{\mathrm{b}}}\right)
=\sqrt{2} \sum_{q, \omega}    (i q)  
f^*(q, \omega) \varphi^\mathrm{q}(q, \omega)
\ee
with $\mathcal{L}_{f,\varphi}\propto f(x,t)\partial_{x} \varphi(x,t)$ and where, in the first identity, we write the Keldysh action in terms of the 
fields $\varphi^{\mathrm{f/b}}$ and, in the second one, in terms of 
quantum component $\varphi^{\mathrm{q}}$ only. This result is standard in the Keldysh formalism and comes directly from the fact that 
a purely classical external force couples \textit{only} with the quantum component $\varphi^{\mathrm{q}}$ of the field.\cite{Kamenev09}
The total $1/f$ effective Keldysh action\cite{Rammerbook} $\mathcal{S}_{1/f}$ for the bosonic field $ \varphi$ derives
from the functional integration 
\be
e^{i\mathcal{S}_{1/f}}=\int\!\! Df\  e^{-\frac{1}{2} \sum_{q, \omega} K^{-1}(q, \omega)|f(q, \omega)|^2}\ e^{i\mathcal{S}_{f,\varphi}}
\ee
which averages on the disorder realizations of the noise potential $f(x,t)$.
 The averaged effective Keldysh action $\mathcal{S}_{1/f}$ can be written in a form similar to Eq.(\ref{KeldyshAction}) with kernel\cite{DallaTorre10} 
\be
\label{G1f}
\mathcal{G}_{1/f}^{-1}(q, \omega)=\left[
\begin{array}{cccc}
0&0\\
0&+ 2 i q^{2}F/|\omega|
\end{array}
\right].
\ee
Here, only the Keldysh q-q component is different from zero. This is a direct consequence of the fact that $1/f$ noise brings the system 
out-of-equilibrium. 
In this case the usual relations between retarded, advanced and Keldysh GFs, dictated by the fluctuation-dissipation 
theorem, are no more valid.

The system, under the external driving force ($1/f$ noise), will reach a stationary condition only in presence of a dissipative mechanism 
that drains the energy accumulated in the system.
Various mechanisms  may introduce dissipation in the edge states.\cite{Cazalilla06,Sassetti94,CastroNeto97,Safi04,Shytov98, Levitov01} Here we limit 
to consider the most general assumption with a dissipative term, induced by the external bath, 
generalizing the Caldeira-Leggett approach to the $\chi$LL.\cite{Caldeira83, CastroNeto97}
 The one dimensional edge mode can be coupled with oscillators through
the current density $j\propto\partial_t\varphi$ or the charge density $\rho\propto\partial_x\varphi$.
Hereafter, we will discuss  the Keldysh action for a generic spectral function of the bath. Later on 
we will focus only on the ohmic behavior.
The general Lagrangian density, which couples edge and harmonic oscillator modes, is
$\mathcal{L}_{\xi, \varphi}\propto \xi (x, \textbf{x}_{\perp}=0,t) \partial_\mu \varphi(x,t) $
where $\mu=t$ ($\mu=x$) describes the coupling with the current (charge) density.
 The field $\xi(x,\mathbf{x}_\perp,t)$ represents a bath of oscillators with extra spatial degrees of freedom $\mathbf{x}_{\perp}$ orthogonal 
 to the 1D system.\cite{CastroNeto97}
 
 Integrating out the harmonic bath degrees of 
 freedom it is easy to obtain the usual Matsubara euclidean effective action\cite{Kamenev09} ($\beta=(k_B T)^{-1}$)
\be
S^{\mathrm{E}}_{\mathrm{diss}}=\frac{1}{2\beta}\sum_{q, i\omega_n}  \mathcal{D}^{-1}(q, i\omega_n) |\varphi(q, i\omega_n)|^2
\label{Mats}
\ee
with $\omega_n=2\pi n/\beta$. The spectral function $\mathcal{D}^{-1}(q, i\omega_n)$ encodes all the dynamical information about the external 
bath and the coupling mechanism.\cite{CastroNeto97} In the Keldysh contour formalism this dissipative 
contribution $\mathcal{S}_{\mathrm{diss}}$
 can be written as in Eq.(\ref{KeldyshAction}) with kernel \cite{Kamenev09}  
\be
\mathcal{G}_{\mathrm{diss}}^{-1}(q, \omega)=
\left[
\begin{array}{cccc}
0&[\mathcal{D}^{A}(q, \omega)]^{-1}\\\
[\mathcal{D}^{R}(q, \omega)]^{-1}& [\mathcal{D}^{-1}(q, \omega)]^K
\end{array}
\right],
\label{Gdiss}
\ee
$[\mathcal{D}^{R/A}(\omega,q)]^{-1}$ being the retarded/advanced analytic continuation of the spectral function $\mathcal{D}^{-1}(q, i\omega_n)$.
 In this case the Keldysh component of the dissipation is computed by using the fluctuation-dissipation theorem\cite{Kamenev09} 
\be
[\mathcal{D}^{-1}]^K=([\mathcal{D}^{R}]^{-1}-[\mathcal{D}^{A}]^{-1})\mathrm{coth}(\beta\omega/2)
\ee
that must be satisfied by a bath in thermal equilibrium.\cite{Kamenev09, DallaTorre10}

As stated before, in this paper we will only consider a specific type of dissipation, the ohmic one. The form of the bath spectral function 
for such case is $\mathcal{D}^{-1}(q, i\omega_n)=\gamma|\omega_n|$, with $\gamma$ the friction coefficient. 
At zero temperature the Keldysh kernel becomes
\be
\label{Gdiss}
\mathcal{G}_{\mathrm{diss}}^{-1}(q, \omega)=\left[
\begin{array}{cccc}
0&-i\gamma\omega\\
+i\gamma\omega&+2i\gamma |\omega|
\end{array}
\right].
\ee

Finally, the total Keldysh action is 
$\mathcal{S}_{\mathrm{tot}}=\mathcal{S}_{0}+\mathcal{S}_{1/f}+\mathcal{S}_{\mathrm{diss}}$
 with total kernel for $\varphi$  
\be
\mathcal{G}^{-1}=\mathcal{G}_{0}^{-1}+\mathcal{G}_{1/f}^{-1}+\mathcal{G}_{\mathrm{diss}}^{-1}
\ee
(cf. Eq.(\ref{chiLLGF}), Eq.(\ref{G1f}) and Eq.(\ref{Gdiss})).
Inverting the kernel, the non-equilibrium Keldysh GFs at zero temperatures read 
\be
\label{Gtot}
\frac{\mathcal{G}}{2\pi\nu}=\left[
\begin{array}{cccc}
-\frac{2 i \tilde{\gamma} |\omega| (1+ (\tilde{F}/\tilde{\gamma}) q^2/\omega^2)}{q^{2}(\omega- vq)^{2} +\tilde{\gamma}^{2} \omega^{2}} &(q(\omega- vq)+ i \tilde{\gamma} \omega)^{-1}\\
(q(\omega- vq) - i \tilde{\gamma} \omega)^{-1}&0
\end{array}
\right]
\ee
where $\tilde{F}=2\pi\nu F$ is the rescaled strength of the noise and $\tilde{\gamma}=2\pi\nu\gamma$ the friction coefficient. The regularization 
factor $\epsilon$ in Eq.({\ref{chiLLGF}) is suppressed because the causal structure is already guaranteed by the dissipative contribution.\cite{Note2} 

It is worth to underline that both the $1/f$ noise and the dissipation terms are relevant perturbations in the Renormalization Group (RG) sense with 
massive coupling constants, namely $\mathrm{dim}[\tilde{F}]=\mathrm{dim}[\tilde{\gamma}]=1$ ($\mathrm{dim}[...] $ indicates the canonical 
mass dimension).
The relevance of these terms will completely spoil the scale invariance property that characterize the standard $\chi$LL theory 
for the edge states. Fortunately, one can demonstrate that for a noisy environment only \emph{weakly} coupled with the 
edge, i.e. $\tilde{\gamma}, \tilde{F}\rightarrow 0$, but with the ratio $\tilde{F}/\tilde{\gamma}$ constant \cite{DallaTorre10} the scale invariance is 
preserved. Indeed in this case, the combined action of the two environmental effects leads only to a {\emph {marginal}} 
perturbation of the 
theory. Consequently the conductance in the Hall bar will be quantized. Indeed, in the limit discussed before, 
with $\tilde{\gamma}\to0$, it is easy to verify that the retarded (advanced) GF $G^R(t)$ ($G^A(t)$), namely the 
anti-transform of the off-diagonal entry 
$\mathcal{G}_{\mathrm{cl,q}}$ ($\mathcal{G}_{\mathrm{q,cl}}$) of the matrix in Eq.(\ref{Gtot}), coincides with the results obtained from the 
retarded (advanced) GFs of the 
free theory\cite{Wen95,Wenbook} given Eq.(\ref{ChiralRetardedGF}). Therefore the linear response shows that a weakly coupled noisy 
environment does not modify the 
conductance of the system with respect to the free $\chi$LL theory.

From the Keldysh GF  $G^K=(\mathcal{G})_{\mathrm{cl,cl}}$ (anti-transform in time of the left-top entry of Eq.(\ref{Gtot})), we can define 
the bosonic correlation function $\tilde{G}^{K}(t)=G^{K}(t)-G^{K}(0)$ with
\be
\tilde{G}^{K}(t)=i\nu g \ln \left[1+ \omega_{\mathrm{c}}^2t^2\right]
\label{GK}
\ee
where $\omega_{\mathrm{c}}=v/a$, with $a$ a finite length cut-off, and 
\be
g=\left(1+ \frac{\tilde{F}}{v^2\tilde{\gamma}}\right).
\label{g}
\ee
Comparing Eq.(\ref{GK}) with the same 
quantity calculated from the free $\chi$LL theory described in Eq.(\ref{chiLLGF}), i.e. 
$\tilde{G}_0^{K}(t)= i\nu \ln \left[1+ \omega_{\mathrm{c}}^2t^2\right]$, we see that the functional dependence remains exactly the same, but 
with an additional renormalization factor $g$.

\subsection{Scaling dimension renormalizations} 
The above result leads to extremely important physical consequences. As a remarkable example we can consider a generic 
$m$-agglomerate quasiparticle (qp) annihilation operator in the bosonized form \cite{Wen95}
\be
\Psi^{(m)}(x)= \frac{e^{i m \varphi(x)}}{\sqrt{2\pi a}}
\ee 
  and the two point greater/lesser GFs
\be
\label{gtrlessGF}
\mathsf{C}_{m}^{>}(t)=\langle \Psi^{(m)}(t)\Psi^{(m)\dagger}(0)\rangle=-\mathsf{C}_{m}^{<}(-t).\\
\ee
These quantities determine the tunneling densities of states of edges and consequently the 
transport properties in a QPC geometry. They can be expressed in terms of the bosonic correlation 
function \cite{Gutman10} $\tilde{G}^{>}(t)=G^{>}(t)-G^{>}(0)$ with\cite{Rammerbook} $G^{>}(t)=(G^{K}(t)+G^{R}(t)-G^{A}(t))/2$ and retarded, 
advanced and Keldysh GFs
obtained from Eq.(\ref{Gtot}). At zero temperature one then has\cite{Gutman10,Mitra11}
\be
\label{greaterGF}
\mathsf{C}_{m}^{>}(g,t)=e^{im^2\tilde{G}^{>}(t)}=
\left[\frac{1}{1+\omega_c^2 t^2}\right]^{\frac{m^2\nu g}{2}} e^{-im^2\nu\phi(t)}
\ee
where 
\be
\phi(t)=\tan^{-1}\left[\frac{\omega_c t}{\sqrt{1+\omega_c^2 t^2}}\right]\underset{\omega_c\to\infty}\to\frac{\pi}{2}\mathrm{sgn}(t). 
\ee
From the comparison of the previous expressions with the results obtained for the free $\chi$LL, 
it is possible to see that the renormalization factor $g$ only influences  
the absolute value of the GF. We explicitly indicate the peculiar functional dependence on $g$ in left hand term of Eq.(\ref{greaterGF}).
The phase instead, as expected,  is not affected being related to the universal statistical properties of the excitations.

We define now the scaling dimension $\Delta(m)$  
of the $m$-agglomerate operator  $\Psi^{(m)}(x)$ as the long-time behavior of the 
two-point GF $|\mathsf{C}^{\gtrless}_{m}(t)|\underset{t\to\infty}\approx|t|^{-2\Delta(m)}$.
This quantity is
\be
\label{ScalingLaughlin}
\Delta(m)=g\Delta_0(m)=g\nu\frac{m^{2}}{2},
\ee
note that the scaling of the raw theory $\Delta_0(m)$ is renormalized by the factor $g$. 
This result induces a modification of the power-law behavior of the transport properties, with respect to the free (unrenormalized) case.

For simplicity we calculate only the single quasiparticle (single-qp) contribution to the back-scattering current, the most dominant one in the Laughlin 
sequence, for the weak-backscattering regime. Notice that, for the Laughlin model, the renormalization mechanism cannot affect the relevance 
of the excitations. We will see that for the models with composite edges this won't be in general the case.
 
We model the QPC in terms of a local tunneling term at $x=0$ between the right- ($R$) and left-($L$) moving  edges 
such as $H_T=\mathbf{t} \Psi_R^{(1)} \Psi_L^{(1)\dagger}+\mathrm{H.c}$.\cite{Sassetti94,Cuniberti96,Braggio01,Cavaliere04} We also assume that 
the edge are affected by different environments and consequently they may have different renormalization parameters $g_{R/L}$ for the right-moving 
edge ($R$) and the left-moving one ($L$).

 The average current at zero temperature, at the lowest order in the tunneling, reads ($\hbar=1$)
\be
\label{IB}
\langle I_B\rangle=e^*\left(\frac{|\mathbf{t}|}{2\pi a}\right)^2\int_{-\infty}^{+\infty}\!\!\!d t\  e^{i E t} \mathsf{C}^>(g_R,t)\mathsf{C}^<(g_L,-t)
\ee
where $E=e^* V$ is the energy involved in the tunneling, with $V$ the bias and $e^*=\nu e$ the single-qp charge. 
From Eq.(\ref{greaterGF}) one has
\be
\mathsf{C}^>(g_R,t)\mathsf{C}^<(g_L,-t)=\left(\frac{1}{1-i\omega_c t}\right)^{\nu(\bar{g}-1)}\left(\frac{1}{1+i\omega_c t}\right)^{\nu(\bar{g}+1)}
\ee  
with $\bar{g}=(g_R+g_L)/2$. From this result one can calculate the expression 
of the current at zero temperature in  Eq.(\ref{IB}) obtaining 
\be
\langle I_B\rangle=e^*\theta(E)\left(\frac{|\mathbf{t}|^2}{a^2\omega_c}\right)\frac{(E/\omega_c)^{2\nu\bar{g}-1}}{\Gamma[\nu( \bar{g}-1)]
\Gamma[\nu(\bar{g}+1)]}\  \mathcal{N}
\ee 
where $\Gamma[x]$ is the Gamma function and $\mathcal{N}=\ _2F_1[\nu(\bar{g}-1),1-\nu(\bar{g}+1),1+\nu(\bar{g}-1),-1]$ is a constant with 
$_2F_1[a,b,c,z]$ the hypergeometric function.\cite{Chang03}

The power-law behavior of the back-scattering current at zero temperature is therefore  $\langle I_B\rangle\propto V^{2\nu \bar{g}-1}$ 
with  renormalized exponent $\nu \bar{g} $.
In the following we will always assume that the renormalization phenomena affects identically the right and left edges.
A generalization to the case of different couplings can be done straightforwardly.

Note that in Eq.(\ref{g}) the strength of the renormalization can take any value $g\geq 1$. The 
same formula suggests that, for fixed environmental contribution ($\tilde{F}/\tilde{\gamma}$ constant), the renormalization 
would be typically stronger for slow propagating modes, due to the explicit dependence on the inverse of the squared mode velocity 
in the expression. 
This renormalization could reach also high values with important modifications of the power-law behavior of transport 
properties.\cite{Chung03, Roddaro03, Roddaro04}

As we have mentioned before, other mechanisms could explain the same renormalization of the 
exponents.\cite{Rosenow02,Khlebnikov06,Yang03,Mandal02,Papa04,Lal08,Overbosch09,Chevallier10} 
However, some of these 
mechanism (such as the 
coupling with phonons) contains intrinsic limitation on the strength renormalization, differently from our 
model where the only real limitation is the request that $g \geq 1$. We will see, in the next sections, that this model can be simply  
generalized to more complex fractional quantum Hall states, such as composite edge states and, more importantly, it also reveals robust 
to the presence of disorder.

\section{Composite edges: Jain sequence}
\label{Jainseq}
\subsection{Model}
\label{modelJain}
Here, we focus on the effects of the out-of-equilibrium noise source in the case of multichannel edge states. The prototype of these Hall states is 
represented by the Jain sequence \cite{Jain89} with filling factor $\nu = p/ (2n p + 1)$, being $n \in \mathbb{N}$ and $p \in \mathbb{Z}$.
Following the hierarchical construction \cite{Wenbook}, one has one charged bosonic mode, analogous to the one described for the Laughlin 
sequence, and $|p|-1$ additional neutral  modes which propagate either  in the same direction ($p > 0$) or in opposite one ($p < 0$). For simplicity 
we restrict the discussion only to the case of two edge modes ($|p|=2$), underlying the differences 
between co-propagating modes ($p > 0$, $\nu = 2/5$) and counter-propagating ones ($p < 0$, $\nu = 2/3$).\cite{Wenbook}
The edge states in the former case are described in terms of two co-propagating bosonic charged fields with different filling 
factors $\nu_1 = 1/3$ and $\nu_2 = 1/15$, such as $\nu= \nu_1 + \nu_2 = 2/5$, while in the latter case the bosonic fields, with $\nu_1 = 1$ and 
$\nu_2 = 1/3$ respectively, propagate in opposite directions leading to $\nu= \nu_1 - \nu_2 = 2/3$. 

The Lagrangian densities are 
\be
\mathcal{L}_\zeta= \sum_{j=1,2}\frac{-1}{4 \pi \nu_j} \partial_{x} \varphi_{j} \left((\zeta)^{j+1}\partial_{t} \varphi_{j}+v_{j} \partial_{x} \varphi_{j}\right) 
\label{Phi0}
\ee
where $\zeta = \pm$ indicates the co-propagating ($\zeta=+$) or counter-propagating ($\zeta=-$) case, and $v_1$, $v_2$ are the velocities 
of the modes and effectively contain the information on intra-edge interactions. The fields commutation 
relations are $[\varphi_j(x),\varphi_k(x')]=i\delta_{jk} \eta_k\nu_k\mathrm{sgn}(x-x')$
where $\eta_k=(\zeta)^{k+1}$ is related to the direction of propagation of the fields ($j,k = 1, 2$).

The two modes are close to each other and interact via the density-density coupling (inter-edge interaction)
\be
\label{coupling}
\mathcal{L}_{12}=\frac{v_{12}}{2\pi\nu_{12}} \partial_{x} \varphi_{1} \partial_{x} \varphi_{2},
\ee
with strength $v_{12}$, where $\nu_{12}=\sqrt{\nu_1 \nu_2}$. 
Notice that, $v_1$, $v_2$ and $v_{12}$ are non-universal parameters related to the intra- and inter-channel interaction strengths. 
Under the reasonable assumption that the confinement potential is sufficiently smooth the two modes are localized in slightly different 
positions. Therefore one reasonably assumes that they effectively "feel" different noisy environments. Indeed,  in general, the trapped charges in 
the substrate beneath the Hall bar affect the $\varphi_1$ and $\varphi_2$ modes in a different way. One can introduce two distinct
$1/f$ noise fields $f_{1/2}(x,t)$ operating on the edge, with noise strength $F_{1/2}$ and spectrum $K_{1/2}(\omega,q)=F_{1/2}/|\omega|$ 
respectively. We will also consider two different ohmic dissipations with friction coefficients $\gamma_{1/2}$. Also in this case the total Keldysh action 
can be written in a form analogous to Eq.(\ref{KeldyshAction}), but in terms of the four components vector 
$\Phi=(\varphi_1^{\mathrm{cl}},\varphi_2^{\mathrm{cl}}, \varphi_1^{\mathrm{q}}, \varphi_2^{\mathrm{q}})^T$, due to the presence of the two fields $\varphi_{1/2}$.
The Keldysh kernel now reads
\begin{widetext}
\be
\label{KernelComposite}
\mathcal{G}^{-1}_{\zeta}=\frac{1}{2\pi}\left[
\begin{array}{cccc}
0&0&\nu_{1}^{-1} \left[q\left(\omega-v_{1} q\right)-i \tilde{\gamma}_1 \omega\right]&\nu_{12}^{-1} v_{12} q^{2}\\
0&0&\nu_{12}^{-1} v_{12} q^{2} &\nu_{2}^{-1}\left[ q\left(\zeta\omega-v_{2}q\right)-i \tilde{\gamma}_2 \omega\right]\\
\nu_{1}^{-1} \left[q\left(\omega-v_{1} q\right)+i \tilde{\gamma}_1 \omega\right] &\nu_{12}^{-1} v_{12} q^{2} & +2 i \nu_{1}^{-1} (\tilde{\gamma}_1 |\omega| +\tilde{F}_1 q^2/ |\omega|)&0\\
\nu_{12}^{-1} v_{12} q^{2}&\nu_{2}^{-1}\left[q\left(\zeta\omega-v_{2}q\right) +i \tilde{\gamma}_2 \omega\right]&0&+2 i \nu_2^{-1}(\tilde{\gamma}_2 |\omega|+\tilde{F}_2 q^2/|\omega|) 
\end{array}
\right]
\ee
\end{widetext}
where  $\tilde{\gamma}_{i}=2\pi\nu_{i}\gamma_{i}$ and  $\tilde{F}_{i}=2\pi\nu_{i}F_{i}$ with $i=1,2$.

\subsection{Interaction effects}
We will discuss now how the presence of a noisy environment could affect 
the renormalization of the $\chi$LL exponents in composite edge systems.
 We will start this analysis by considering the presence of  interaction between the channels described in Eq.(\ref{coupling}). Let us start 
 to discuss the case of co-propagating channels such as 
$\nu=2/5$. For such cases, indeed, the clean system, i.e. with no static disorder along the edge, properly describes 
the physics of the edge states. In the next subsection we will investigate the renormalization effects for $\nu=2/3$ (counter-propagating modes)  
assuming instead the presence of the static disorder, which is crucial in the equilibration process between counter-propagating modes.

The Keldysh action kernel for $\nu=2/5$ is given by $\mathcal{G}^{-1}_{+}$ in Eq.(\ref{KernelComposite}) with $\nu_1 =1/3$ and $\nu_2=1/15$. 
To better analyze the problem it is useful to make a rescaling $\varSigma$  and a 
rotation $\mathcal{R}(\theta)$ of the fields $\varphi_{1/2}$. In the new basis $\varphi'_{1/2}$ we have 
\be
\label{rotation}
\binom{\varphi_1}{\varphi_2}
=\underbrace{\left[\begin{array}{cc}\sqrt{\nu_1}&0\\0&\sqrt{\nu_2}\end{array}\right]}_{\varSigma}\cdot\underbrace{\left[\begin{array}{cc}\cos(\theta)&\sin(\theta)\\-\sin(\theta)&\cos(\theta)\end{array}\right]}_{\mathcal{R}(\theta)}
\cdot\binom{\varphi'_1}{\varphi'_2}
\ee
where the angle $\theta$ satisfies $\tan(2\theta)=2v_{12}/(v_1-v_2)$. The new fields eigenmodes $\varphi'_{1/2}$ are 
decoupled, with respect to the density-density interaction, and have different velocities 
\be
\label{newvelocities}
v'_{1,2}=\left(\frac{v_1+v_2\pm\sqrt{(v_1-v_2)^{2}+4v_{12}^2}}{2}\right).
\ee 
From the above relations we get a criterium, called \emph{stability}, which requires these velocities to be always positive.\cite{Kane95}
This is reflected in a constraint  $v_{12}^2\leq v_1 v_2$ between the intra- and inter-mode couplings.\\ 
At the same time the dissipative and $1/f$ terms acquire off-diagonal contributions due to the transformation in Eq.(\ref{rotation}). 
In particular, we can see what happen to those terms by focusing at
the $2\times2$ bottom left block-matrix, that coincides with the q-cl component of the Kernel in Eq.(\ref{KernelComposite}). 
In the new $\varphi'$ basis it becomes
\be
(\mathcal{G}^{\prime -1})_{\mathrm{q,cl}}
=
\frac{q}{2\pi}\left[\begin{array}{cc} \omega-v'_{1} q & 
0  
\\ 
0 
& \omega-v'_{2}q \end{array}\right]
+(\mathcal{G}'^{-1}_{\mathrm{diss}})_{\mathrm{q,cl}}
\ee
where the dissipative contribution is no more diagonal and reads
\be
\label{rotgamma}
(\mathcal{G}'^{-1}_{\mathrm{diss}})_{\mathrm{q,cl}}=
\mathcal{R}^{T}(\theta)\cdot\left[\begin{array}{cc} i \tilde{\gamma}_1\omega &0\\0&i \tilde{\gamma}_2\omega\end{array}\right]\cdot\mathcal{R}(\theta)
\ee
with the rotation $\mathcal{R}(\theta)$ defined in Eq.(\ref{rotation}).

Notice that the advanced component ($2\times2$ top right block-matrix) $(\mathcal{G}^{\prime -1})_{\mathrm{cl,q}}$ can be easily 
derived from the previous result by 
complex conjugation, $(\mathcal{G}^{\prime -1}(\omega,q))_{\mathrm{cl,q}}=\left((\mathcal{G}^{\prime -1}(\omega,q))_{\mathrm{q,cl}}\right)^*$.
The Keldysh component $(\mathcal{G}'^{-1})_{\mathrm{q,q}}$ of Eq.(\ref{KernelComposite}) ($2\times 2$  bottom right 
block-matrix) transforms in the new basis  according to Eq.(\ref{rotgamma})
\be
\label{1fprime}
\frac{(\mathcal{G}'^{-1})_{q,q}}{2i|\omega|}=
\mathcal{R}^{T}(\theta)\cdot\left[\begin{array}{cc} 
\tilde{\gamma}_1 +\tilde{F}_2 \frac{q^2}{|\omega|^2}&0\\0&
\tilde{\gamma}_2+\tilde{F}_2 \frac{q^2}{|\omega|^2}
\end{array}\right]\cdot\mathcal{R}(\theta)
\ee
where the linear dependence on the coefficients  
$\tilde{F}_1$, $\tilde{F}_2$ is now explicit.

In the limit of weak contribution of noise and dissipation (see Sec.\ref{laughlin}), i.e. $\tilde{\gamma}_{i},\tilde{F}_{i}\to0$ but keeping the ratios 
$\tilde{F}_{i}/\tilde{\gamma}_{i}$ constant, it is possible to calculate all the Keldysh GFs following the same approach used for the Laughlin case.\\ 
To simplify the discussion we consider only the case where the effective friction coefficients of the dissipative contributions are the 
same, $\tilde{\gamma}_1=\tilde{\gamma}_2=\tilde{\gamma}$, allowing only different strengths for the $1/f$ noise.\cite{Note5} This assumption 
greatly simplify our discussion without affecting the key results.

For $\tilde{\gamma} \to 0$ one recovers again the standard form of the retarded/advanced GFs for composite edges. A direct consequence of 
this fact is that the argument discussed in Ref.~\onlinecite{Kane95}, where the conductance of a multichannel edge of interacting 
co-propagating modes is calculated using the retarded GFs, is still valid here. Therefore, also for the multichannel edge the Hall bar current is 
correctly quantized $I=\nu \mathfrak{g}_0 V_H$, independent of the value of the inter-edge 
interaction $v_{12}$.\cite{Note6}

The presence of an external noisy environment modifies the scaling dimension of the excitations. Now we analyze this point, 
taking care of the presence of inter-edge coupling $v_{12}$. In particular 
we will show that the scaling properties are no more universal and depend, in general, on inter-edge coupling. This result differentiates 
from the standard theory \cite{Kane95} that predicts only universal scaling properties in the co-propagating case. It is important to stress 
that this is a \emph{direct consequence} of the presence of a 
noisy external environment. To better demonstrate this fact, we will consider now all the possible excitations and their scaling dimensions.

In the bosonized form a generic excitation is written in terms of a linear combination of the bosonic field $\varphi_{1/2}$ as
\be
\label{Psialpha}
\Psi^{(\alpha_1,\alpha_2)}(x)\propto e^{i\left[\alpha_1\varphi_1(x)+\alpha_2\varphi_2(x)\right]}
\ee
with $\alpha_{1/2}$ coefficients that determine the considered excitation.\cite{Ferraro08, Ferraro10a, Ferraro10c} Here, we just recall that 
the charges of all the excitations are integer multiples of the fundamental single qp, e.g. for $\nu=2/5$, the fundamental charge is  
$e^* = e/5$ ($e$ the 
electron charge). The statistical properties of the excitations are directly connected to the values of $\alpha_{1}$ and $\alpha_2$ 
and the commutation 
relations of the $\varphi_{1/2}$ fields\cite{Ferraro10a,Ferraro10c}, while the scaling dimensions depends additionally, as we will see, on the 
presence of a noisy environment. The two-point correlation function of the operator is\cite{Gutman10}
\be
\label{CompositeGreaterGF}
\mathsf{C}^>_{\alpha_1,\alpha_2}(t)=\langle \Psi^{(\alpha_1,\alpha_2)}(t) {\Psi^{(\alpha_1,\alpha_2)}}^\dagger(0)\rangle=e^{\sum_{i,j=1,2}\alpha_i\tilde{G}^>_{ij}(t)\alpha_j}
\ee
where, in the second equality, we introduced the greater GFs $\tilde{G}^>_{jk}(t)=G^>_{jk}(t)-G^>_{jk}(0)$ such that 
$G^>_{jk}(t)=-i\langle\varphi_j(t)\varphi_k(0)\rangle$ for the $\varphi_j$ fields with $j,k=1,2$.

In the new basis $\varphi'_{1/2}$, taking the limit of weak coupling with the environment $\tilde{\gamma},\tilde{F}_1,\tilde{F}_2\to 0$, the 
Keldysh GFs read ($j,k=1,2$) 
\be
\label{Gkvarphiprime}
\tilde{G}'^K_{jk}(t)=i\delta_{jk}g'_j \ln [1+\omega'^2_{c,j}t^2]
\ee
where the cut-offs are $\omega'_{c,j}=v'_j/a$ with $j=1,2$. The "mixed" terms $\tilde{G}'^K_{jk}(t)$ with $j\neq k$ vanish due to the 
assumption $\tilde{\gamma}_1=\tilde{\gamma}_2=\tilde{\gamma}$ on the dissipative friction coefficients. The renormalization 
coefficients for the two normal modes 
are now
\be
\label{gprime}
g'_j=\left(1+\frac{\tilde{F}_+-(-)^{j}\tilde{F}_-\cos(2\theta)}{2 v_j'^2 \tilde{\gamma}}\right)
\ee
with $\tilde{F}_\pm=\tilde{F}_1\pm\tilde{F}_2$ and the mode velocities $v_j'^2$ given in Eq.(\ref{newvelocities}). 
Notice that the renormalization parameters depend on  the coupling strength $v_{12}$ and the
mode velocities, through the angle $\theta$. This appears a quite natural generalization of the result given in Eq.(\ref{g}).
From this result one can calculate $\tilde{G}'^>_{jk}(t)$ with the same procedure used for the Laughlin sequence, in the 
form of Eq.(\ref{greaterGF}), after the proper replacement of the renormalization parameter $\nu g \to g'_1,g'_2$ and of the
cut-offs $\omega_c \to v'_{1}/a,v'_{2}/a$. The greater GF in Eq.(\ref{CompositeGreaterGF}) can be expressed as
\be
\tilde{\mathbf{G}}^>(t)=\varSigma\cdot\mathcal{R}(\theta)\cdot\tilde{\mathbf{G}}'^>(t)\cdot\mathcal{R}^T(\theta)\cdot\varSigma
\ee
where we used the compact matrix notation with $\tilde{G}^>_{jk}(t)=(\tilde{\mathbf{G}}^>(t))_{jk}$ with the rescaling matrix $\Sigma$ and 
the rotation matrix $\mathcal{R}(\theta)$ defined in Eq.(\ref{rotation}). 

From the long-time behavior of the two-point correlation function of Eq.(\ref{CompositeGreaterGF}) one can calculate the scaling
 dimension $\Delta^{(\alpha_1,\alpha_2)}$ of the $\Psi^{(\alpha_1,\alpha_2)}(x)$ operator
\beq
\Delta^{(\alpha_1,\alpha_2)}=&\frac{1}{2}\left\{\nu_1\  \alpha_1^2\left[g'_1\cos^2(\theta)+g'_2\sin^2(\theta)\right]\nonumber\right.\\
&+\nu_2\  \alpha_2^2\left[g'_2\cos^2(\theta)+g'_1\sin^2(\theta)\right]
\nonumber
\\
&\left.+\sqrt{\nu_1\nu_2}\ \alpha_1\alpha_2\sin(2\theta)\left[g'_1-g'_2\right]\right\}
\eeq
where we see an explicit dependence on the coupling $v_{12}$ and the mode velocities $v_{1,2}$, via the angle $\theta$. Note that, in the 
absence of environmental 
effects ($g'_{1}=g'_{2}=1$), the scaling dimensions reduce to the standard 
$\Delta_0^{(\alpha_1,\alpha_2)}=(\nu_1\alpha_1^2+\nu_2\alpha_2^2)/2$ obtained for hierarchical theories.
Furthermore, if the renormalizations of the normal modes are exactly the same $g'_1=g'_2=g$ the scaling dimension become independent of the 
angle $\theta$ (i.e. the coupling $v_{12}$) with $\Delta^{(\alpha_1,\alpha_2)}=g\Delta_0^{(\alpha_1,\alpha_2)}$.

 In conclusion, we showed that the scaling of a generic excitation in presence of a noisy environment is influenced by the strength of the 
 coupling $v_{12}$ even for co-propagating modes. This strongly differs from 
 the standard result\cite{Kane95} where the 
 scaling is \emph{independent} from the coupling. This fact  shows that scaling dimension in the presence of $1/f$ noise are, in general, \emph{no 
 more universal}, i.e. determined only by the coefficients  $(\alpha_1,\alpha_2)$, the filling factor $\nu$ and the model of composite edge we consider. 
 As a remarkable consequence, in the presence of environmental effects, the relevance between the excitations could differ from 
 the raw theory. We have already discussed this phenomenology in relations to experimental observations.\cite{Ferraro08,Ferraro10a}}

We will see now that the most important advantage of the presented model is its robustness with respect to static 
impurity disorder along the edge. Indeed, all the results up to now are 
essentially based on the assumption of a clean edge, without any contribution coming from static disorder. We will also see that including such 
contribution the role of inter-mode interactions will be less important but the scaling dimensions will be still affected by renormalizations 
effects due to the noisy environment. 

\subsection{Disorder effects}
A more realistic discussion of edge states in real samples requires the inclusion of static disorder along the edge.\cite{Note4} 
The disorder plays a fundamental role to recover the proper quantization of the Hall 
conductance for counter-propagating modes. We refer the reader to the seminal paper 
by Kane and Fisher in Ref.~\onlinecite{Kane95} for a detailed discussion of the disorder dominated phase in the hierarchical theories. 

In the following we will analyze the case of $\nu=2/3$ where the two channels are counter-propagating. The discussions about the disorder 
effects in the composite edge, presented here, can be also generalized 
to the whole Jain sequence. In the next section we adapt the
argument even to the $\nu=5/2$ state.

The Keldysh action for the multichannel edge state at $\nu=2/3$ under the influence of $1/f$ noise and dissipation is given by the kernel 
$\mathcal{G}_-^{-1}$ of Eq.(\ref{KernelComposite}) with $\nu_1=1$ and $\nu_2=1/3$. The effect of static disorder on the $\chi$LL channels  
can be naturally described by adding two more terms to the action. 

The first one describes the coupling of two static disorder potential profiles $V_i(x)$ 
with the charge densities $\rho_i(x)=\partial_x\varphi_i(x)/(2\pi)$ of the two channel $\varphi_i$ composing the edge
$\nu=2/3$ with $i=1,2$. The Lagrangian of this term, which affects locally the two channels is $\mathcal{L}_{V, \varphi}\propto \sum_{i}V_i(x)\partial_{x} \varphi_i(x)$. These forward scattering terms can be easily eliminated from the action 
by a simple redefinition of the $\varphi_i(x)$ fields and will be neglected in the following.\cite{Kane95} 
  
The second term describes the effect of the disorder in terms of impurity scattering, i.e. how the disorder potential mediates 
the electron transfer between the two 
counter-propagating modes. These tunneling terms have the main effect to equilibrate the two channels when they are at different potentials restoring the proper value of the quantized conductance.\cite{Kane95} 
This random tunneling term is
\be
\mathcal{L}_{\rm{rdm}}= \xi(x) e^{i\left[\varphi_{1}(x)+3\varphi_{2}(x)\right]}+h.c.,
\label{Lagrangian_random}
\ee
with $\xi(x)$ a complex random tunneling amplitude. This process leads to the destruction of an electron into the $\nu=1$ channel ($e^{i\varphi_1(x)}$) and its creation  ($e^{i3\varphi_{2}(x)}$) into the $\nu=1/3$ one and viceversa.\cite{Kane95, Kane94} For simplicity it is assumed $\xi(x)$ as a Gaussian random variable $\delta$-correlated in space satisfying
\be
\label{disorderLagrangian}
\langle\xi^{*}(x)\xi(y)\rangle_{\rm{ens}}=W \delta(x-y),
\ee
where $\langle...\rangle_{\rm{ens}}$ indicates the ensemble average over the realizations of disorder. 

To further analyze the disorder terms, it is convenient to express the system in the basis $\varphi'_{1/2}(x)$
that diagonalizes the problem with respect to inter-edge coupling. We follow essentially the same 
procedure used for the $\nu=2/5$ case. The transformation now is given by the composition of a rescaling $\varSigma$ and a Lorentz boost 
$\mathcal{B}(\chi)$ with rapidity $\chi$, instead of the standard rotation used for co-propagating modes.
The relation between the old fields $\varphi_{1,2}$ with the new ones $\varphi'_{1,2}$ is
\be
\label{Lorentzboost}
\binom{\varphi_1}{\varphi_2}
=\underbrace{\left[\begin{array}{cc}\sqrt{\nu_1}&0\\0&\sqrt{\nu_2}\end{array}\right]}_{\varSigma}\cdot\underbrace{\left[\begin{array}{cc}\cosh(\chi)&\sinh(\chi)\\\sinh(\chi)&\cosh(\chi)\end{array}\right]}_{\mathcal{B}(\chi)}
\cdot\binom{\varphi'_1}{\varphi'_2}
\ee
 where $\tanh(2\chi)=-2v_{12}/(v_1+v_2)$.\\ 
 Note that we can calculate the scaling dimension for a generic qp operator $\Psi^{(\alpha_1,\alpha_2)}(x)$ defined in Eq.(\ref{Psialpha})
 following the same steps as in the previous section. It is also useful, without loosing generality, to assume $\tilde{\gamma}_1=\tilde{\gamma}_2=\tilde{\gamma}$ but keeping as free parameters the strengths of the $1/f$ noise $\tilde{F}_{1/2}$. 
 We find the scaling dimension
 \beq
 \Delta^{(\alpha_1,\alpha_2)}=&\frac{1}{2}\left\{\nu_1\  \alpha_1^2\left[g'_1\cosh^2(\chi)+g'_2\sinh^2(\chi)\right]\right.\nonumber\\
&+\nu_2\  \alpha_2^2\left[g'_2\cosh^2(\chi)+g'_1\sinh^2(\chi)\right]
\nonumber
\\
&\left.+\sqrt{\nu_1\nu_2}\ \alpha_1\alpha_2\sinh(2\chi)\left[g'_1+g'_2\right]\right\}
\label{Deltaa1a2}
\eeq
where the contribution of the Lorentz boost is explicit in the terms $\cosh^2(\chi)$ and $\sinh^2(\chi)$. The renormalization factors of the new modes $g'_i$ are now ($j=1,2$)
\be
\label{gboost}
g'_j=\left(1+\frac{\tilde{F}_+-(-)^{j}\tilde{F}_- \mathrm{sech}(2\chi)}{2 v_j'^2 \tilde{\gamma}}\right),
\ee
where the eigenmode velocities $v'_i$ are in Eq.(\ref{newvelocities}) and $\tilde{F}_\pm$ are defined after Eq.(\ref{gprime}). 
Note that a direct comparison with the co-propagating expression shows that, in the counter-propagating case, the $\mathrm{sech}(2\chi)$ takes the 
role of the $\cos(2\theta)$ of Eq.(\ref{gprime}), but the form of the renormalization factors remains essentially the same.  

We address now the role of the disorder terms, starting with the inspection of the scaling $\Delta_{\mathcal{O}}=\Delta^{(1,3)}$ of the tunneling 
disorder operator  $\mathcal{O}\propto e^{i\left[\varphi_{1}(x)+3\varphi_{2}(x)\right]}$ introduced in Eq.(\ref{Lagrangian_random}).
One can demonstrate that the flow equation for the inter-edge disorder strength $W$ in Eq.(\ref{disorderLagrangian}) 
is \cite{Kane94, Kane95, Giamarchi88}
\be
\label{RGflow}
\frac{d W}{d l}=(3-2 \Delta_{\mathcal{O}})W.
\ee
In particular, if $\Delta_{\mathcal{O}}<3/2$ the disorder is a relevant contribution driving the system in the so called disorder-dominated phase. 
For such phase the Hall bar conductance $\mathfrak{g}$ is universal (i.e. independent of the environment and the intra- and 
inter-mode couplings) and is properly quantized at the value $\mathfrak{g}=\nu \mathfrak{g}_0$.\cite{Kane94}  
All the discussions of quantum Hall states with counter-propagating modes are typically done assuming the system being exactly in this phase. \\ 
On the contrary, if the scaling of the electron tunneling is $\Delta_{\mathcal{O}}>3/2$, the disorder is irrelevant and, at the fixed 
point, the conductance is no more universal depending on the intra- and inter-edge interactions.\cite{Kane94}

In general the environmental
 effects, through the parameters 
$\tilde{F}_{i}/\tilde{\gamma}_{i}$ and the intra- and inter-mode couplings, could affect the scaling $\Delta_{\mathcal{O}}$ making the 
discussion quite cumbersome. As a simple check, we first need to recover the standard result obtained in absence of any environmental effects, 
namely the ratios $\tilde{F}_1/\tilde{\gamma},\tilde{F}_2/\tilde{\gamma}\to 0$.  From Eq.(\ref{Deltaa1a2}) and the definition of the 
operator $\mathcal{O}$ one gets 
\be
\label{Delta023}
\Delta^0_{\mathcal{O}}=\lim_{\tilde{F}_i\to 0}\Delta^{(1,3)}=2\frac{v_1+v_2-\sqrt{3} v_{12}}{\sqrt{(v_1+v_2)^2-4 v^2_{12}}}
\ee
which coincides with the result of Kane, Fisher and Polchinski.\cite{Kane94}
\begin{figure}
\centering 
\includegraphics[width=0.48\textwidth,]{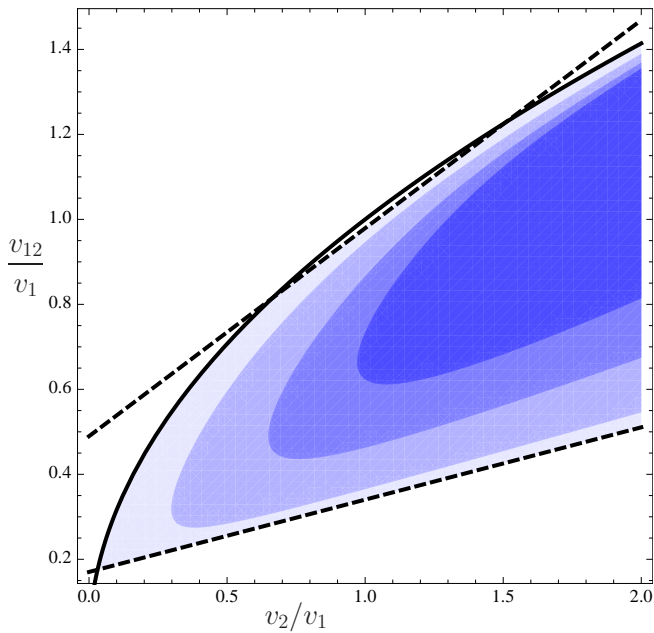}
\caption{Colored areas represent the disorder dominated phase for $\nu=2/3$ in the parameters space $(v_2/v_1,v_{12}/v_1)$ for different 
strength of the noise. In light blue the case without the noisy environment \cite{Kane94}, i.e. $\Delta^0_{\mathcal{O}}<3/2$ (see in the text). 
This area is limited by the \emph{stability} criterium (solid line) and the 
two dashed lines representing respectively the maximum/minimum value of the ratio $v_{12}/v_1$ to get the disordered phase  
$\Delta^0_{\mathcal{O}}<3/2$. 
Other colored areas represent $\Delta_{\mathcal{O}}<3/2$ with successive reduction of the disordered phase due to the increasing of noise 
strength $\tilde{F}_1/(v^2_1\tilde{\gamma})=0$ (lighter blue), $ 0.1, 0.5, 1$ (darker blue) with $\tilde{F}_2=0$. }   
\label{fig:Fig1}
\end{figure}
It is convenient now to measure the velocities in units of $v_1$. For the values of the couplings where the \emph{stability} criterium 
$v_{12}^2\leq v_1 v_2$ is satisfied 
we can determine when $\Delta^0_{\mathcal{O}}<3/2$. In Fig.~\ref{fig:Fig1} we report, in the plane 
$(v_2/v_1,v_{12}/v_1)$, the regions where this condition is fulfilled. The area is delimited by the \emph{stability} (solid black) curve and 
two (dashed black) lines representing respectively  the maximum/minimum value of $v_{12}/v_1$ compatible with the disorder dominated phase. These 
two lines are given by $v_{12}/v_1=(4\sqrt{3}/21\pm\sqrt{5}/14)(1+v_2/v_1)$.\\
Now we can evaluate how this area changes under the presence of a noisy environment. One could expect, in analogy with the 
renormalizations induced by coupling with phonons\cite{Rosenow02}, that the effects of an external environment 
lead always to an enhancement of the scaling dimension 
$\Delta_{\mathcal{O}}\geq \Delta^0_{\mathcal{O}}$. 
A direct consequence of this fact would be the progressive reduction of the region of existence of the disorder dominated phase. This is 
explicit in the figure 
where we calculated the regions where $\Delta_{\mathcal{O}}<3/2$, using Eq.(\ref{Deltaa1a2})
varying $\tilde{F}_i/(v_1^2\tilde{\gamma})=0, 0.1, 0.2, 0.3$ for a fixed  ratio $\tilde{F}_2/(v_1^2\tilde{\gamma})=0$. Notice that for very strong 
noise the disordered dominated phase could be completely washed out. 

We conclude this discussion observing that, for moderate noise strength, the disorder dominated phase is still present even if the condition 
on the inter- and intra-mode coupling are modified. 
Our analysis generalizes some of the results of Ref.~\onlinecite{Kane94} in presence of a noisy environment.\\
The robustness of the proposed model for the renormalization of the exponents in 
presence of disorder along the edge is one of the most important result of this paper.

We will discuss now quantitatively how renormalizations are affected by noise intensity. In the disorder dominated phase the system 
naturally decouples in charged and neutral contributions.\cite{Kane94, Kane95} Therefore, it is convenient to change the basis from 
the original $\varphi_{1/2}$ 
to the charged $\varphi_\rho$ and neutral fields $\varphi_{\sigma}$
\be
\binom{\varphi_{\rho}}{\varphi_{\sigma}}=\frac{1}{\sqrt{2}}\binom{\sqrt{3}(\varphi_{1}+\varphi_{2})}{3\varphi_{2}+\varphi_{1}}
\label{charged_neutral}
\ee
as obtained from the transformation in Eq.(\ref{Lorentzboost}) with $\tanh(\chi^*)=-\sqrt{1/3}$.\cite{Note7}
The action is expressed in the form of Eq.(\ref{KernelComposite}), but with the propagation velocities  $v_i$ with index $i=\rho,\sigma,\rho\sigma$ given by
\be
\left[\begin{array}{cc}v_\rho&v_{\rho\sigma}\\v_{\rho\sigma}&v_\sigma\end{array}\right]=\mathcal{B}(\chi^*)^T\cdot\left[\begin{array}{cc}v_1&v_{12}\\v_{12}&v_2\end{array}\right]\cdot\mathcal{B}(\chi^*)
\ee
with $\mathcal{B}(\chi^*)$ the Lorentz boost in Eq.(\ref{Lorentzboost}). The same transformation defines the noise strengths $\tilde{F}_i$, in the 
new basis as
\be
\label{Frhosigma}
\left[\begin{array}{cc}\tilde{F}_\rho&\tilde{F}_{\rho\sigma}\\\tilde{F}_{\rho\sigma}&\tilde{F}_\sigma\end{array}\right]
=\frac{1}{2}\left[\begin{array}{cc}3\tilde{F}_1+\tilde{F}_2&\sqrt{3}(\tilde{F}_1+\tilde{F}_2)\\\sqrt{3}(\tilde{F}_1+\tilde{F}_2) &\tilde{F}_1+3\tilde{F}_2\end{array}\right]
\ee
in terms of the coefficients $\tilde{F}_{1/2}$ of Eq.(\ref{KernelComposite}). An equivalent transformation can be written for the friction 
coefficients of the dissipative bath, with the introduction of the quantities
$\tilde{\gamma}_{\rho},\tilde{\gamma}_\sigma,\tilde{\gamma}_{\rho\sigma}$, as linear combinations of $\tilde{\gamma}_{1/2}$.

The off-diagonal terms of the action containing 
$v_{\rho\sigma}$ are irrelevant in the RG sense and can be neglected at the fixed point. This was clearly shown in 
Ref.~\onlinecite{Kane95}. In the limit of weak coupling such that $\tilde{\gamma}_i,\tilde{F}_i\to0$, but keeping constant 
the ratios $\tilde{F}_i/\tilde{\gamma}_i$, with $i=1,2$, the environmental contributions are marginal in the RG sense.\cite{DallaTorre11} 
This shows that, at the fixed point of the disorder 
dominated phase, we could safely neglect the residual coupling between charged and neutral modes but we \emph{have to} include the 
noisy environmental contributions.  In the following we will take 
$v_{\rho\sigma}=0$ keeping explicitly the dissipative and $1/f$ noise terms in account.  

We observe that, in the case of the coupling with 1D phonon $\varphi_{\mathrm{ph}}$ modes\cite{Rosenow02}, one has also a
 term analogous to the one proportional to $v_{\rho \sigma}$ discussed above. The canonical mass dimension of the phonons in $1+1$ dimensions is the same 
 of a chiral bosonic field $\mathrm{dim}[\varphi_{\mathrm{ph}}]=\mathrm{dim}[\varphi_{1,2}]$. As a natural consequence,  this coupling term 
 becomes RG irrelevant in the disordered dominated fixed point as already discussed. This shows that, even 
if the coupling with phonons could in principle generate renormalizations of the scaling exponent, in the disorder dominated phase 
the phonons are effectively decoupled from the system and their renormalization effects do not survive against disorder. This indicates that our 
model is qualitatively different and  
presents concrete advantages in comparison with other mechanisms especially for all those cases where counter-propagating modes are 
present and, consequently, the disorder dominated phase has to be considered.

We can now evaluate the GFs along the same line followed in the previous section for 
$\nu=2/5$. Also in this case we assume $\tilde{\gamma}_1=\tilde{\gamma}_2=\tilde{\gamma}$ and consider the strengths of the $1/f$ noises $\tilde{F}_{1/2}$ as free parameters.

In the limit of $\tilde{\gamma}\to 0$ the retarded/advanced GFs are exactly the same as the ones in Ref.~\onlinecite{Kane95} and consequently 
the edge conductance returns the appropriate 
quantized value of $\mathfrak{g}=\nu \mathfrak{g}_0$. 
For the Keldysh GF contributions of  the charged and neutral fields we get, in the disordered phase, a result identical to Eq.(\ref{Gkvarphiprime}), 
where the only non-zero GF are the 
$\tilde{G}^K_{\rho\rho}$ and $\tilde{G}^K_{\sigma\sigma}$. These are characterized by the cut-off energies $\omega_i=v_i/a$ 
with $i=\rho,\sigma$ and by
renormalization parameters
\be
\label{renormalizNC}
g_{\rho}=\left(1+\frac{\tilde{F}_{\rho}}{2v^2_\rho\tilde{\gamma}}\right)\qquad g_{\sigma}=
\left(1+\frac{\tilde{F}_{\sigma}}{2v^2_\sigma\tilde{\gamma}}\right).
\ee
They coincide with Eq.(\ref{gboost}) choosing $\chi=\chi^*$ and using the definition of Eq.(\ref{Frhosigma}).
\begin{figure}
\centering 
\includegraphics[width=0.48\textwidth]{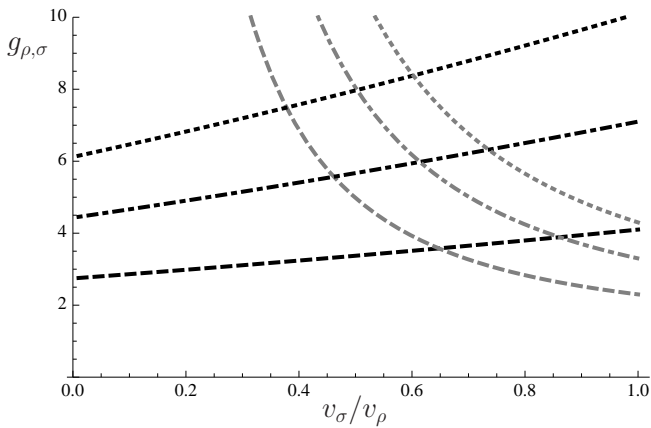}
\caption{Renormalization parameters $g_\rho$ (black curves) and $g_\sigma$ (gray curves) as a function of the ratio $v_\sigma/v_\rho$. Different 
linestyles correspond to different noise strength $\tilde{F}_1/(v_1^2 \tilde{\gamma})=1$ (dashed), $2$ (dot-dashed), $3$ (dotted) having kept fixed 
the value of $\tilde{F}_2/(v_1^2 \tilde{\gamma})=0.1$}
\label{fig:Fig2}
\end{figure}
Their values depend on the noise strength $\tilde{F}_i$, the dissipation $\tilde{\gamma}$ and 
on the neutral and charged mode velocities  $v_i$. The Fig.~\ref{fig:Fig2} shows how the charged renormalization parameter (black curves) and the 
neutral one (gray curves) depend on the ratio $v_\sigma/v_\rho$  for different values of the $1/f$ noise strengths. At fixed noise 
strength and increasing the ratio $v_\sigma/v_\rho$ the charged mode renormalization rises while the 
neutral one decreases. 
This behavior is directly connected to the dependence on the inverse of the squared mode velocities of Eq.(\ref{renormalizNC}). 
When the modes velocities become small the renormalization parameters increase rapidly.

Interestingly, it is also possible to obtain the counterintuitive condition $g_\sigma>g_\rho$. Following physical intuition indeed, it appears 
natural to assume that neutral bosonic modes are less coupled with the environment with respect to the charged ones. 
Nevertheless, this intuition failsbecause  the neutral bosonic modes, in the composite edges, derives from a particle-hole combination 
between the two 
modes and are strongly affected by the differences in the noisy environments (differential mode) where the charge 
modes instead are affected 
by the common mode only.

In conclusion the dependence of the renormalization parameters on the noise strengths 
$\tilde{F}_{1/2}$ (see Eq.(\ref{Frhosigma})) guarantees the possibility to get very high renormalization values for 
almost any values of the velocities ratio $v_\rho/v_\sigma$.  Note that these high values are sometimes 
necessary to fully explain the experimental observations.\cite{Ferraro10b}

We conclude this section showing that, rescaling the two fields $\varphi_c=\sqrt{2/3}\varphi_\rho$ and  $\varphi_n=\sqrt{2}\varphi_\sigma$, 
it is possible to write all qp operators as\cite{Kane94, Ferraro10b}
\be
\Psi^{(m,l)}\propto e^{i[ (m/2) \varphi_{c}+(l/2) \varphi_{n}]},
\ee
with the coefficients $m,l\in \mathbb{Z}$ and with the same parity. These operators destroy an $m$-agglomerate, namely an excitation 
with charge $m e^{*}$ being $e^{*}=e/3$ the minimal charge allowed by the model. Their scaling dimensions become
\be
\Delta(m,l)=
\frac{1}{2}\left[\left(\frac{2}{3}\right) g_{\rho}\left(\frac{m}{2}\right)^2 +2g_{\sigma} \left(\frac{l}{2}\right)^{2}\right],
\ee
where the $g_\rho$ and $g_\sigma$ renormalize the charge and neutral sectors of the excitation separately. Obviously we recover the scaling 
dimension reported in the literature \cite{Wen95, Ferraro10b} for $g_{\rho}=g_{\sigma}=1$. The last formula shows that, in the 
disorder dominated phase, the presence of a noisy environment naturally leads to different 
 renormalizations for the neutral and charged modes. Consequence of this fact is the possibility to change the relevance of the
 excitations and, indeed when $g_\rho,g_\sigma\neq1$, this could happen due to environmental effects we are discussing. This phenomenology 
 could have a deep impact on transport properties of the QPC especially in the 
 weak backscattering regime where the dominant excitations are different from the electrons. The possibilities opened by our model 
 for composite edges could therefore explain the extremely rich phenomenology observed in QPC transport at low temperatures 
 for these systems. In Ref.~\onlinecite{Ferraro10a} and 
 Ref.~\onlinecite{Ferraro10b} 
 we have discussed in detail the experiments on noise and transport in QPC for $\nu=2/3$. To fully match the theory with the data the presence of 
 the renormalization parameters $g_\rho, g_\sigma\geq1$ was sufficient. Here, we have shown that a noisy environment can be considered a 
 proper renormalization mechanisms, robust to unavoidable disorder effects.

\section{Composite edges: the $\nu=5/2$ case}
\label{fivehalf}
\subsection{Anti-Pfaffian model}
Another relevant example of composite edge state is represented by $\nu=5/2$. Possible descriptions have been proposed for this state 
predicting both Abelian \cite{Halperin93} and non-Abelian \cite{Moore91, Fendley07, Lee07, Levin07} statistical properties for the elementary 
excitations. Particularly interesting is the so called anti-Pfaffian model \cite{Lee07, Levin07}, supporting non-Abelian statistics, 
that seems to be indicated by experimental 
evidences as a proper description for this state.\cite{Radu08, Bid10} According to this model, the edge states are described as a narrow 
region at $\nu=3$ with nearby a Pfaffian edge of holes with $\nu=1/2$.\cite{Lee07} Assuming the second Landau level as the ``vacuum'', the edge is 
modeled in terms of a single $\nu=1$ bosonic branch $\varphi_1$ and a counter-propagating $\nu=1/2$ Pfaffian branch \cite{Fendley07}, composed 
by a bosonic mode $\varphi_2$ and a Majorana fermion $\bpsi$. 
 
 The Lagrangian for the free system is $\mathcal{L}_0=\mathcal{L}_-+\mathcal{L}_{12}+\mathcal{L}_{\psi}$
 where the bosonic contribution $\mathcal{L}_-$ and $\mathcal{L}_{12}$ are given in Eq.(\ref{Phi0}) and Eq.(\ref{coupling}) respectively 
 with $\nu_1=1$ and $\nu_2=1/2$. The Lagrangian describing the free evolution of the Majorana fermion \bpsi\   in the Ising sector is
 \be
 \mathcal{L}_{\psi}= i \bpsi \left( -\partial_{t} +v_{\psi} \partial_{x}\right) \bpsi,
 \ee
 with propagation velocity $v_{\psi}$. In addition to the free theory we have the coupling of the bosonic modes $\varphi_{1/2}$ with the 
 different noisy environments ($1/f$ noise and the dissipative ohmic bath) surrounding them. Also in this case we consider coupling with 
 the $1/f$ noise strengths $\tilde{F}_i$ and friction coefficients of the dissipative baths $\tilde{\gamma}_1=\tilde{\gamma}_2=\tilde{\gamma}$. Note that the noise 
 and the dissipation couple electrostatically with $\varphi_{1/2}$ but not with the neutral Ising sector of the theory that is decoupled 
 from electromagnetic environment.  The total Keldysh bosonic action coupled with the noisy environment has the  
 kernel $\mathcal{G}_-^{-1}$ of Eq.(\ref{KernelComposite})  with $\nu_1=1$ and $\nu_2=1/2$. The Lagrangian density is completed  
 by the addition of the disorder term 
\be
\mathcal{L}_{rdm}= \xi(x) \bpsi(x)\  e^{i \left[\varphi_{1}(x)+2 \varphi_{2}(x)\right]}+h.c.
\ee
that describes the random electron tunneling processes 
which equilibrate the two branches, in fully analogy with $\nu=2/3$.
The complex random coefficients $\xi(x)$, Gaussian distributed, satisfy also Eq.(\ref{disorderLagrangian}).
This unavoidable contribution guarantees that the appropriate value of the Hall resistance is recovered in the disorder dominated phase. 
The RG flow equation for the disorder term $W$ is the same of Eq.(\ref{RGflow}), with $\Delta_{\mathcal{O}}$ the scaling dimension 
of the tunneling operator $\mathcal{O}\propto \bpsi e^{i \left[\varphi_{1}+2 \varphi_{2}\right]}$. Consequently, investigating when 
$\Delta_{\mathcal{O}}< 3/2$, it identifies the conditions for a disorder dominated phase of $\nu=5/2$.\cite{Lee07, Levin07}

We firstly identify the conditions of the existence of the disorder 
dominated phase as a function of the couplings $v_i$ and the noisy environment.  The 
scaling is 
\be
\Delta_{\mathcal{O}}=\frac{1}{2}+\Delta^{(1,2)},
\ee
where the first term in the sum represents the contribution of the Ising sector (Majorana fermion) and the second one 
is the bosonic contribution of Eq.(\ref{Deltaa1a2}) with $\nu_1=1$ and $\nu_2=1/2$. The bosonic contribution can 
be indeed derived 
following exactly the same steps considered for $\nu=2/3$. 
\begin{figure}
\centering 
\includegraphics[width=0.48\textwidth]{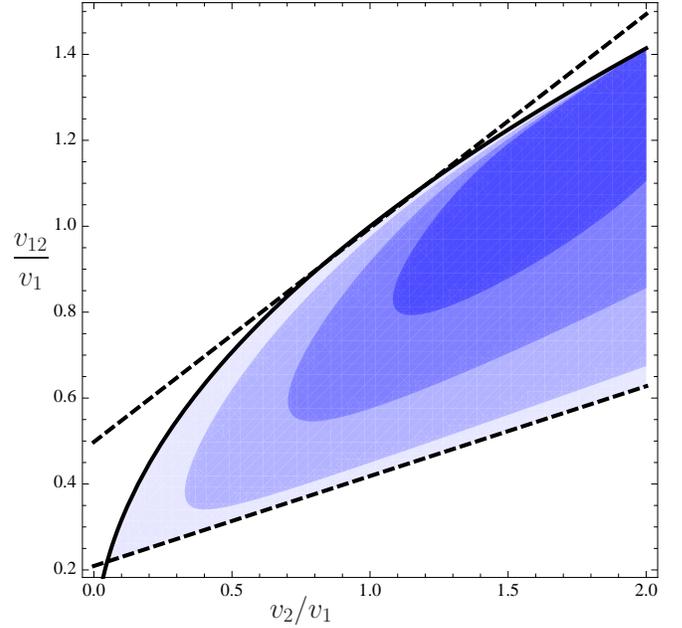}
\caption{Colored areas represent the disorder dominated phase for $\nu=5/2$ in the 
parameter space $(v_2/v_1,v_{12}/v_1)$ for different strength of the noise. In light blue the case without the noisy 
environment \cite{Kane94}, $\Delta^0_{\mathcal{O}}<3/2$ (see in the text). 
This area is limited by the \emph{stability} condition (solid line) and the 
two dashed lines representing respectively the maximum/minimum value of the ratio $v_{12}/v_1$ to get the disordered phase. Other colored area 
represents the successive reduction of the disordered phase due the increasing of noise 
strength $\tilde{F}_1/v^2_1\tilde{\gamma}= 0$(lighter blue), $0.1, 0.5, 1$(darker blue) keeping fixed $\tilde{F}_2=0$.}
\label{fig:Fig3}
\end{figure}
The scaling dimension in general depends on the renormalization 
parameters $g'_{i}$, defined in Eq.(\ref{gboost}).
Without the noisy environment $g'_{1}=g'_{2}=1$ we then recover 
\be
\label{Delta052}
\Delta^0_{\mathcal{O}}=\lim_{\tilde{F}_i\to0} \Delta_{\mathcal{O}}=\frac{1}{2}\left(1+\frac{3 (v_1+v_2)-4\sqrt{2} v_{12}}{\sqrt{(v_1+v_2)^2-4 v^2_{12}}}
\right)
\ee
that is the scaling dimension of the intra-edge electron tunneling reported in the literature.\cite{Levin07} 

The region of existence of the disorder dominated phase ($\Delta_{\mathcal{O}}< 3/2$) is represented in Fig.~\ref{fig:Fig3} for 
different velocities of the modes
$(v_2/v_1,v_{12}/v_1)$. The lines delimiting the area are the same discussed in the previous section: the 
\emph{stability} condition (black solid curve) and the two lines (dashed black) that limit the values of $v_{12}/v_1$, i.e. 
$v_{12}/v_1=(3\sqrt{2}\pm\sqrt{3})(1+v_2/v_1)/12$. 

The discussion hereafter goes in parallel with what we have done for $\nu=2/3$. The noisy 
environment will further restrict the set of values of intra- and inter-mode couplings where 
the system is dominated by the disordered phase. In 
the figure this is represented by the progressive reduction of the  colored area: from the lighter blue to the darker blue when  the 
environmental noise increases. If the noise 
becomes strong enough the disorder dominated phase could even disappear. 

Also for $\nu=5/2$, at the fixed point of the disorder dominated phase, the system naturally decouples into a charged bosonic 
mode $\varphi_{c}=\varphi_{1}+\varphi_{2}$  with velocity $v_{\mathrm{\rho}}$ and a neutral counter-propagating sector (one
 bosonic mode $\varphi_{n}=\varphi_{1}+2 \varphi_{2}$ and one Majorana fermion $\bpsi$ with the same velocity 
 $v_{\mathrm{\sigma}}$).\cite{Levin07,Lee07} It is again natural 
 to introduce the charged $g_\rho$ and neutral $g_\sigma$ renormalization parameters, according to Eq. (\ref{renormalizNC}). 
The renormalizations can be very strong, for realistic values of the ratio $v_\sigma/v_\rho$, and satisfying the condition $g_\sigma>g_\rho$ 
as we anticipated for  $\nu=2/3$. In conclusion our model explains the values of the renormalizations proposed in Ref.~\onlinecite{Carrega11} for $\nu=5/2$.\cite{Dolev10}
 
\subsection{Agglomerate dominance}
Here, we will discuss the effects of noisy environment on the relevance of excitations in the anti-Pfaffian model.
Using the charged and neutral modes basis one 
 can express the more general qp operator as \cite{Levin07, Carrega11} 
\be
\Psi^{(\mbox{\scriptsize \boldmath $\chi$},m,l)}\propto \mbox{ \boldmath $\chi$}(x) e^{i \left[(m/2) \varphi_{c}+(l/2) \varphi_{n}\right]},
\ee
where the integer coefficients $m,l$ and the Ising field operator \mbox{ \boldmath $\chi$}$(x)$ define the admissible excitations. In the Ising sector  
\mbox{ \boldmath $\chi$}$(x)$ can be $\mathbf{I}$ 
(identity operator), \mbox{ \boldmath $\psi$}$(x)$ (Majorana fermion) or \mbox{ \boldmath $\sigma$}$(x)$ (spin operator). 
The monodromy condition force $m$, $l$ to be even integers for \mbox{ \boldmath $\chi$}$=\mathbf{I},$\mbox{\boldmath $\psi$} and odd integers 
for \mbox{ \boldmath $\chi$}$=$\mbox{ \boldmath $\sigma$}. The charge associated to the above operator is $(m/4)e$ depending on the 
charged mode contribution only, while its scaling dimension is 
\be
\Delta(\mbox{\boldmath $\chi$}, m, l)=\frac{1}{2}\delta_{\mbox{\scriptsize \boldmath $\chi$}}+\frac{g_{\mathrm{\rho}}}{16} m^{2}+\frac{g_{\sigma}}{8}l^{2}\,,
\label{Delta}
\ee 
with $\delta_{\mathbf{I}}=0$, $\delta_{\mbox{\scriptsize \boldmath $\psi$}}=1$ and $\delta_{\mbox{\scriptsize \boldmath $\sigma$}}=1/8$. 
Note that, as stated before, the contribution of the Ising sector to the scaling 
dimension is not affected by any renormalization.

We adopted the previous formula to predict the scaling dimension and the transport properties in the
experiment done by the Heiblum group at Weizmann.\cite{Carrega11}. We found a 
good agreement with the experiment where, at the lowest temperatures, the dominant excitation is the $2$-agglomerate $2 e^*=e/2$, that is 
described by the operator $\Psi^{(\mathbf{I},2,0)}$. 
Our explanation clarifies why the anomalous 
increasing of the effective charge is observed at extremely low temperatures. 

\begin{figure}
\centering 
\includegraphics[width=0.49\textwidth]{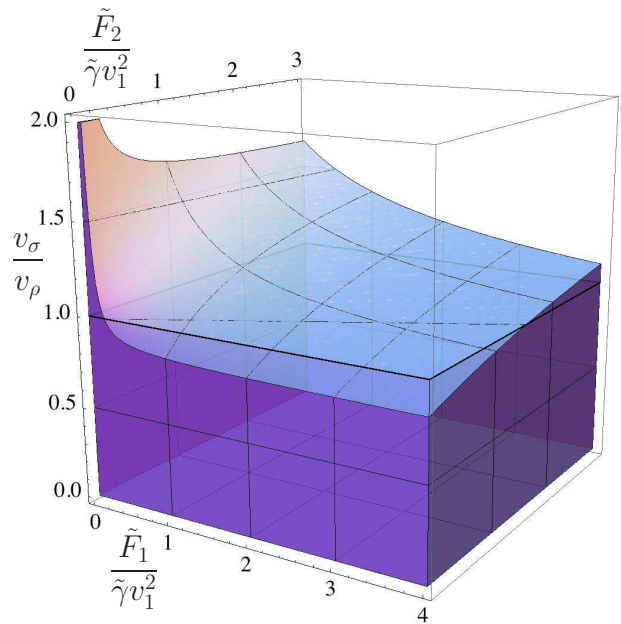}
\caption{Three dimensional picture of the region (colored) where the inequality $g_\rho<(1+2g_\sigma)/3$ is fulfilled, namely where the $2$-agglomerate dominates with respect to the single-qp. On the vertical axis is reported the ratio $v_{\sigma}/v_{\rho}$, while in the plane the renormalization factors $\tilde{F}_1/(\tilde{\gamma}v_1^2)$ and $\tilde{F}_2/(\tilde{\gamma}v_1^2)$. The plane $v_{\sigma}/v_{\rho}=1$ is highlighted with a thick line.}
\label{fig:Fig4}
\end{figure}

Let's see now when the noise environmental parameters determine the dominance of the $2$-agglomerate. In general,
the excitation with the lowest scaling dimension dominates the properties in the low energy sector. 
Without any renormalization ($g_\rho=g_\sigma=1$) the scaling dimensions 
are exactly the same: $\Delta(\mathbf{I}, 2, 0)=\Delta(\mbox{\boldmath $\sigma$}, 1, \pm1)=1/4$.
So \emph{only} the presence of environmental renormalization will determine the dominance of an excitation over the other. 
The effect of renormalizations 
are indeed crucial to make the single-qp excitation - described by the operator $\Psi^{(\mbox{\scriptsize \boldmath $\sigma$},1,\pm1)}$
with charge $e^*=e/4$ - less relevant than the agglomerate.\cite{Note8}

The agglomerate with charge $e/2$ will be 
dominant over the single-qp if $\Delta(\mathbf{I}, 2, 0)<\Delta(\mbox{\boldmath $\sigma$}, 1, \pm1)$ so we get the 
inequalities\cite{Carrega11} 
\be
g_\rho<\frac{1+2g_\sigma}{3}.
\ee 

In Fig.~\ref{fig:Fig4} we show the domain where the agglomerate $\Psi^{(\mathbf{I},2,0)}$ is a dominant 
over the single-qp $\Psi^{(\mbox{\scriptsize \boldmath $\sigma$},1,\pm1)}$. 
We see that agglomerates are more easily dominant for $v_\sigma/v_\rho<1$ - the regime probably valid in the real samples.
Conversely, when $v_\sigma/v_\rho\gtrsim 2$, 
the dominance of agglomerate is possible only at very small values of noise strength as show by the peak in the figure. 

Note that, for small values of $\tilde{F}_2$ and strong enough $\tilde{F}_1$, it is also possible to have the dominance of the 
single-qp for $v_\sigma/v_\rho<1$. 
In the last figure this corresponds to the volume underneath the plane identified by the thick line that coincide with $v_\sigma/v_\rho=1$. 

In conclusion the dominance of the agglomerate is quite common and only in the case of neutral modes velocity similar to the charged modes 
\emph{and} in the presence of a noisy environment the the single-qp could be \emph{more} dominant. Anyway, we want to mention that the 
excitation that dominates at very low 
energy, potentially, couldn't be also the 
dominant at higher energies, i.e. by increasing bias or temperature. This explains why the single-qp seems to be the 
dominant charge carriers in measurements carried out at higher 
temperatures\cite{Radu08,Dolev08}, as we discussed in more details in Ref.~\onlinecite{Carrega11}.

We conclude this section commenting on the need of a correct identification of the dominant excitations at low energy.  
We recall that one of the most important properties of anti-Pfaffian (Pfaffian) states 
for  $\nu=5/2$  is the possibility to support excitations which satisfying non-Abelian statistics. Indeed, the single-qp is represented by the operator $\Psi^{(\mbox{\scriptsize \boldmath $\sigma$},1,\pm1)}$ that, due to the peculiar fusion rule in the Ising sector
\mbox{\boldmath $\sigma$} $\times$ \mbox{\boldmath $\sigma$}$=\mathbf{I}+$\mbox{\boldmath $\psi$} is intrinsically non-Abelian. 
On the other hand the agglomerate is Abelian, being represented in term of the operator $\Psi^{(\mathbf{I},2,0)}$, i.e. with an identity operator $\mathbf{I}$ on the Ising sector. 

Therefore, the dominance of the agglomerates with respect to the single-qp could have important consequence on the real possibility to manipulate 
non-Abelian excitations with the help of QPC setups. The hope to encode topological protected quantum computation protocols in this system 
may be potentially affected by this issue. Counterintuitively, given the previous analysis, a noisy environment could become a helpful resource 
leading, in some regions of the parameter space, to the dominance of the non-Abelian single-qp. 

In perspective we like to mention that our approach and also many of the discussed results could be recovered also 
for other models, such as the Pfaffian or the Abelian 331.\cite{Halperin93,Lin12}  This shows that, for an large class of models of edges states, the renormalization phenomena induced 
by the noisy environment can play an important role influencing the physics in the low energy regime.

\section{Conclusions}
\label{Conclusions}
We have presented a renormalization mechanism of the tunneling exponent in the $\chi$LL theories for edge states, based 
on the joint effects of the weak coupling with out-of-equilibrium $1/f$ noise and dissipation. The model is very general and can be applied 
to many different states, such as in the Jain sequence or even the anti-Pfaffian model for $\nu=5/2$.

Considering the paradigmatic case of the Laughlin sequence, we showed how a noisy environment could modify the Luttinger exponents. 
The direct consequences of this renormalization are derived for the QPC current in the weak-backscattering regime, mainly focusing on 
the effects on the power-law behavior as a function of bias.

In the Jain sequence, and in particular for $\nu=2/5$, we have investigated how the scaling dimensions of the excitations are affected by the interplay between the inter-channel couplings and the noisy environment. Here, the possibility of a change in the dominance of the excitations is reported. 

The rich phenomenology induced by these facts was already considered by us\cite{Ferraro08, Ferraro10a, Ferraro10c}, and we found  
a good match with the experimental observations.\\
The case of counter-propagating modes has been analyzed in detail. We investigated how the noisy environment modifies the conditions of 
stability of the disorder dominated phase. We demonstrated that, for moderate noise strength, the renormalization mechanism is robust 
against disorder and remains valid at the fixed point of the disorder dominated phase. This is a crucial result, because of all the quantum Hall 
edge theories with counter-propagating modes require the presence of static disorder to guarantee the equilibration along the edge and the 
proper universal value of the quantum resistance experimentally observed.
This robustness makes our model a good candidate for a realistic renormalization mechanism of the Luttinger exponent while other models, 
such as the coupling with 1D phonons or other bosonic baths, might not survive in presence of disorder.

In the last part of the paper we discussed the $\nu=5/2$ case considering the non-Abelian anti-Pfaffian model for the edge states. In analogy 
with the previous analysis we studied the effect of external environments and their role on the disorder dominated phase. 

Our proposal for the renormalization mechanism seems to be applicable to a plethora of cases giving a convincing and rather simple unified perspective.
Our results suggest that the values of the Luttinger exponents, that typically in the literature are related to universal features of the adopted theoretical models, have to be taken with care  due to the presence of unavoidable noisy environments that can modify, even consistently, some of the predictions.

\section*{Acknowledgements.} We thank E. G. Dalla Torre for valuable discussions and acknowledge the support of the CNR STM 2010 program and the EU-FP7 via ITN-2008-234970 NANOCTM.


\begin{thebibliography}{10}
\bibitem{Note1b} Note that the coupling strength can be rescaled through an appropriate redefinition of the parameter $F$.
\end{thebibliography}
\end{document}